\documentclass[english,letterpaper,aps,prl,twocolumn,showpacs]{revtex4}
\usepackage{lmodern}
\usepackage[T1]{fontenc}
\usepackage[latin1]{inputenc}
\usepackage{amsmath}
\usepackage{amssymb}
\usepackage{graphicx}

\makeatletter

\newcommand*\LyXThinSpace{\,\hspace{0pt}}

\@ifundefined{textcolor}{}
{%
 \definecolor{BLACK}{gray}{0}
 \definecolor{WHITE}{gray}{1}
 \definecolor{RED}{rgb}{1,0,0}
 \definecolor{GREEN}{rgb}{0,1,0}
 \definecolor{BLUE}{rgb}{0,0,1}
 \definecolor{CYAN}{cmyk}{1,0,0,0}
 \definecolor{MAGENTA}{cmyk}{0,1,0,0}
 \definecolor{YELLOW}{cmyk}{0,0,1,0}
}

\usepackage{lmodern}

\makeatletter

\usepackage{lmodern}

\makeatletter

\usepackage{lmodern}

\makeatletter

\makeatletter

\usepackage{epsfig}

\usepackage{dcolumn}

\usepackage{bm}

\usepackage{bbm}

\usepackage{wasysym}

\usepackage{marvosym}

\usepackage{subfloat}
\usepackage{color}
\usepackage{subfigure}

\usepackage{blindtext}

\newlength{\textwidthm}

\makeatother

\makeatother

\makeatother

\makeatother

\makeatother

\usepackage{babel}
\begin{document}

\title{Superconducting states of Semi-Dirac Fermions at Zero and Finite
Magnetic Field }

\author{Bruno Uchoa$^{*}$ and Kangjun Seo}

\affiliation{Department of Physics and Astronomy, University of Oklahoma, Norman,
OK 73069, USA}
\email{uchoa@ou.edu}

\selectlanguage{english}%

\date{\today}
\begin{abstract}
We address the superconducting singlet state of anisotropic Dirac
fermions that disperse linearly in one direction and parabolically
in the other. For systems that have uniaxial anisotropy, we show that
the electromagnetic response to an external magnetic flux is extremely
anisotropic near the quantum critical point of the superconducting
order. In the quantum critical regime and above a critical magnetic
field, we show that the superconductor may form a novel exotic \emph{smetic}
state, with a stripe pattern of flux domains. 
\end{abstract}


\maketitle
\emph{Introduction.$-$ }Semi-Dirac metals form a class of two dimensional
(2D) systems with chiral quasiparticles that disperse linearly in
one direction and quadratically in a different direction \cite{Banarjee}.
In the presence of spin-orbit coupling, the zero energy crossings
of the Dirac cones remain protected by space group symmetries of the
crystal \cite{Young} and may have a non-zero Chern number \cite{Huang,Saha}.
Examples of semi-Dirac metals include a variety of systems, including
VO$_{2}$/TiO$_{2}$ heterostructures \cite{Pardo,Huang}, and strained
crystals such as graphene and black phosphorus, which can undergo
a topological phase transition towards a semi-Dirac phase \cite{Montambaux,Rodian}.
Semi-Dirac cones have been experimentally realized on the top layer
of black phosphorus under electric field effects, which tune the system
from a trivial band gap insulator to a band inverted system \cite{Kim}. 

In this rapid communication, we explore the properties of $s$-wave
singlet states for semi-Dirac fermions in the vicinity of a quantum
critical point (QCP). We show that semi-Dirac fermion superconductors
have an exotic electromagnetic response to an applied magnetic flux.
Due to the anisotropy of the quasiparticles, the stiffness of the
order parameter to the penetration of a magnetic flux can be highly
anisotropic near the QCP. In that regime, we show that semi-Dirac
metals with uniaxial anisotropy can effectively behave as type I superconductors
along one direction, and as type II superconductors in the other.
As a result, instead of vortices, the system may form a novel smetic
state with stripes of superconducting domains intercalated by thin
normal strips of magnetic flux \cite{note3-1}. 

\begin{figure}[b]
\includegraphics[scale=0.4]{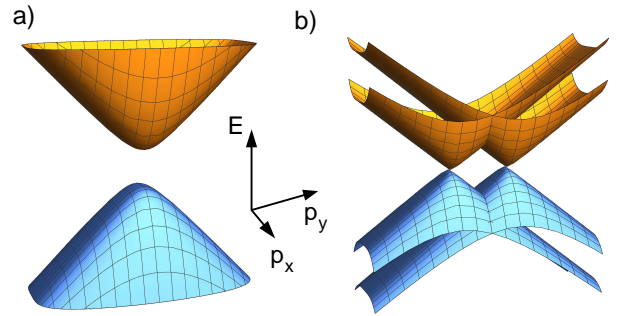}

\caption{{\small{}Energy spectrum of the superconducting singlet states of
semi-Dirac fermions. a) Intra-orbital paring state, which is fully
gapped around each nodal point. b) inter-orbital state, where the
nodes split and remain gapless. The gapped state is dominant. }}
\end{figure}

\emph{Hamiltonian.$-$} For concreteness, we start from a two-orbital
model on a square lattice, 
\begin{equation}
\mathcal{H}_{0}(\mathbf{k})\equiv\mathbf{g}(\mathbf{k})\cdot\vec{\sigma},\label{eq:0}
\end{equation}
where $\mathbf{g}=(g_{x},g_{y},g_{z})$ is a vector with components
$g_{x}(\mathbf{k})=4t^{\prime}(\cos k_{x}-\cos k_{y})^{2}$, $g_{y}(\mathbf{k})=0$
and $g_{z}(\mathbf{k})=2t(\cos k_{x}+\cos k_{y})$, $t$ and $t^{\prime}$
are effective hopping parameters, $k$ is the momentum with respect
to the center of the square Brillouin zone and $\sigma_{x}$ and $\sigma_{z}$
are Pauli matrices in the orbital space \cite{Banarjee}. The low
energy Hamiltonian is described by semi-Dirac fermions around four
nodal points $\mathbf{k}_{0}=(\pm\frac{1}{2},\pm\frac{1}{2})\pi$,
with 
\begin{equation}
\mathcal{H}_{0,\alpha}^{(+)}(\mathbf{p})=\frac{p_{x}^{2}}{2m}\sigma_{x}-\alpha vp_{y}\sigma_{z}\equiv\mathbf{h}_{+,\alpha}(\mathbf{p})\cdot\vec{\sigma},\label{eq:1}
\end{equation}
describing the pair of nodes at $\mathbf{k}_{0}=\alpha(\frac{1}{2},\frac{1}{2})\pi$
($\alpha=\pm$), where $\mathbf{p}$ is the momentum away from the
nodes (we set $\hbar\to1$), with $p_{x}$ and $p_{y}$ as momentum
coordinates along the two diagonal directions ($1,\bar{1}$) and $(1,1)$
respectively. $m$ is the mass of the quasiparticles that disperse
quadratically with momentum $p_{x}$ along one direction and $v$
gives the Fermi velocity of the quasiparticles that disperse linearly
along the perpendicular direction. The other two nodes at $\mathbf{k}_{0}=\alpha(\frac{1}{2},-\frac{1}{2})\pi$
are described by the low energy Hamiltonian
\begin{equation}
\mathcal{H}_{0,\alpha}^{(-)}(\mathbf{p})=-\alpha vp_{x}\sigma_{x}+\frac{p_{y}^{2}}{2m}\sigma_{z}\equiv\mathbf{h}_{-,\alpha}(\mathbf{p})\cdot\vec{\sigma}.\label{eq:1-2}
\end{equation}
In both sets of pairs, opposite nodal points are related by time reversal
symmetry (TRS). 

The Bogoliubov-deGennes Hamiltonian for Eq. (\ref{eq:0}) is
\begin{equation}
\mathcal{H}_{\text{BdG}}(\mathbf{k})=\left(\begin{array}{cc}
\mathcal{H}_{0}(\mathbf{k}) & \hat{\Delta}\\
\hat{\Delta} & -\mathcal{T}\mathcal{H}_{0}(\mathbf{k})\mathcal{T}^{-1}
\end{array}\right),\label{eq:H}
\end{equation}
where the $2\times2$ matrix $\hat{\Delta}$ gives superconducting
order parameter matrix elements in the orbital space and $\mathcal{T}\mathcal{H}_{0}(\mathbf{k})\mathcal{T}^{-1}=\mathcal{H}_{0}(\mathbf{k})$
is the TRS operation of the Hamiltonian. 

In the singlet state, there are two possible pairing channels. The
first one is the intra-orbital pairing state, with pairing matrix
elements $\hat{\Delta}=\Delta\sigma_{0}$, which result in a fully
gapped low energy spectrum
\begin{equation}
\pm E_{\mathbf{p}}=\pm\sqrt{h^{2}(\mathbf{p})+\Delta^{2}},\label{eq:E-1}
\end{equation}
with $h(\mathbf{p})=|\mathbf{h}(\mathbf{p})|$ (the valley indexes
are omitted). The second channel is the inter-orbital pairing state,
$\hat{\Delta}=\Delta\sigma_{x}$, which leads to gapless superconductivity,
$\pm E_{\mathbf{p},s}=\pm\sqrt{h_{x}^{2}(\mathbf{p})+(h_{z}(\mathbf{p})+s\Delta)^{2}}$
with $s=\pm$ indexing two additional branches, shown in Fig. 1b.
For a given attractive interaction, the fully gapped state lowers
the free energy of the system more than the gapless one by pushing
the energy states down towards the bottom of the band, as shown in
Fig. 1a. In this letter, we will focus on the dominant instability
and address the thermodynamic and electromagnetic properties of the
fully gapped state. 

\emph{Critical behavior.$-$} The free energy of the superconducting
state is $F(T)=\Delta^{2}/g-T\sum_{\mathbf{k},\gamma}\log\{2+2\cosh(\gamma E_{\mathbf{k}}/T)\}$,
with $\gamma=\pm$ indexing the particle and hole branches of the
spectrum respectively, $T$ is the temperature and $g>0$ is the effective
attractive interaction that leads to formation of Cooper pairs. In
mean field, minimization of the free energy with respect to $\Delta$
(assumed to be real) gives the standard BCS equation of state $g^{-1}=\sum_{\mathbf{q}}\tanh(\frac{1}{2T}E_{\mathbf{q}})/2E_{\mathbf{q}}.$
Using the parametrization where $h_{x}(\mathbf{p})=p_{x}^{2}/2m=h\cos\theta$
and $h_{z}(\mathbf{p})=vp_{y}=h\sin\theta$, with $\theta\in[-\frac{\pi}{2},\frac{\pi}{2}]$,
the density of states can be written in terms of the Jacobian of the
transformation $(p_{x},p_{y})\to(h,\theta)$ \cite{Adroguer},
\begin{equation}
\rho(h,\theta)=\frac{N_{0}}{8\pi^{2}}\frac{\sqrt{2mh}}{v\cos\theta}.\label{eq:rho2}
\end{equation}
Integration in $\theta$ gives the actual density of states, $\rho(h)=2\int_{-\pi/2}^{\pi/2}\mbox{d}\theta\rho(h,\theta)=\rho_{0}\sqrt{h}$,
where $\rho_{0}=\sqrt{m}N_{0}K(\frac{1}{2})/(\pi^{2}v)$, with $K(\frac{1}{2})\approx1.85$
an elliptic function and $N_{0}$ is the node degeneracy.

At zero temperature and half filling, the phase transition is quantum
critical due to the vanishing DOS at the nodal point \cite{Kotov,Uchoa,Zhao}.
Near the QCP, the mean field zero temperature gap scales with the
coupling as
\begin{equation}
\Delta(0,g)=\frac{1}{(c_{1}\rho_{0})^{2}}\left(\frac{1}{g_{c}}-\frac{1}{g}\right)^{2}\theta(g-g_{c}),\label{eq:Delta}
\end{equation}
where $c_{1}=\Gamma^{2}\!\left(\frac{3}{4}\right)/\sqrt{\pi}\approx0.85$,
with $\Gamma(x)$ a gamma function, and $g_{c}=1/(\sqrt{\Lambda}\rho_{0})$
is the critical coupling defined in terms of the effective energy
bandwidth $\Lambda$. In the gapless state, the critical coupling
is $g_{c}^{\prime}=3/(\sqrt{2\Lambda}\rho_{0})>g_{c}$, and hence
the gapped instability clearly prevails. In the two band model (\ref{eq:0})
where $m^{-1}=16t^{\prime}$, $v=2\sqrt{2}t$ and $\Lambda\sim2t$,
then $g_{c}/t=8\pi^{2}/[N_{0}K(\frac{1}{2})]\sqrt{t^{\prime}/t}$.
In the limit where $t^{\prime}/t\lll1$, the critical coupling can
be small enough to allow the QCP physics to be accessed experimentally.
In general, since $g_{c}\propto v/\sqrt{m}$ scales with the velocity
and mass of the quasiparticles, the critical coupling can be further
lowered with strain effects \cite{Wei}. 

\begin{figure}
\begin{centering}
\includegraphics[scale=0.5]{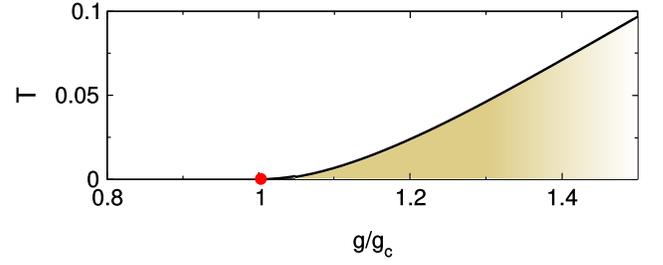}
\par\end{centering}
\caption{{\small{}Phase diagram of temperature (in units of the cut-off $\Lambda$)
vs coupling for the fully gapped state in the vicinity of the QCP
at $g=g_{c}$. The order parameter scales as $\Delta\propto(1-g_{c}/g)^{\beta}$
near the QCP, with $\beta=2$ in mean field.}}
\end{figure}

The mean field critical temperature is given by $T_{c}(g)\approx c_{1}^{2}\Delta(0,g),$
as shown in Fig. 2. In the critical regime, 
\begin{equation}
\Delta(T\approx T_{c},g)\approx2.02\,\Delta(0,g)\sqrt{\frac{T_{c}}{T}-1}.\label{eq:Delta2}
\end{equation}
 The specific heat at fixed volume is defined as $C_{V}=-T\mbox{d}^{2}F/\mbox{d}T^{2}.$
At the phase transition, the specific heat jump normalized by specific
heat in the normal side of the transition, $\delta C_{V}\approx0.71$
\cite{deltaC}. In the case of Dirac fermions in 2D (graphene), $\delta C_{V}\approx0.35$
\cite{uchoa}, while in the Fermi liquid case $\delta C_{V}\approx1.43$
\cite{Tinkham}. 

\emph{Supercurrent.}$-$ To calculate the Meissner response to an
external magnetic flux, we include a vector potential $\mathbf{A}$
in Hamiltonian (\ref{eq:H}) in the Coulomb gauge, explicitly breaking
TRS, \emph{$\mathcal{T}\mathcal{H}_{0}(\mathbf{k}-\frac{e}{c}\mathbf{A})\mathcal{T}^{-1}=\mathcal{H}_{0}(\mathbf{k}+\frac{e}{c}\mathbf{A})$}.
When the Fermi level is at the neutrality point, the energy spectrum
can be calculated analytically, 
\begin{equation}
E_{\mathbf{k},s}(\mathbf{A})=\sqrt{g_{D}^{2}+g_{\xi}^{2}+\Delta^{2}+2s\sqrt{(\mathbf{g}_{D}\!\cdot\!\mathbf{g}_{\xi})^{2}+g_{\xi}^{2}\Delta^{2}}},\label{eq:E}
\end{equation}
with $s=\pm$, and $g_{D,\xi}=|\mathbf{g}_{D,\xi}|$, with $\mathbf{g}_{D,\xi}(\mathbf{k})=\frac{1}{2}\sum_{s=\pm}s^{q}\mathbf{g}(\mathbf{k}-s\frac{e}{c}\mathbf{A})$,
where $q=0,\,1$ describe the symmetric ($D$) and anti-symmetric
($\xi$) combinations in the vector potential, respectively. 

The calculation of the supercurrent from Eq. (\ref{eq:0}) and (\ref{eq:H})
can be done in a very general way for any arbitrary vector $\mathbf{g}=(g_{x},g_{y},g_{z})$
defined in terms of generic functions of momenta $g_{i}(\mathbf{p}),\,i=x,y,z$,
provided TRS is preserved at zero field. From the minimal coupling
between currents and electromagnetic fields, $\mathcal{H}_{\text{I}}=\frac{1}{c}\mathbf{j}\cdot\mathbf{A}$,
the current operator is $\mathbf{j}=c\partial\mathcal{H}_{\text{BdG}}/\partial\mathbf{A}$.
The supercurrent in the London limit is $\langle\mathbf{j}\rangle=-c\,\mbox{tr}\frac{1}{\beta}\sum_{i\omega,\mathbf{k}\in BZ}\left[\partial_{\mathbf{A}}\mathcal{H}_{\text{BdG}}(\mathbf{k},\mathbf{A})\right]\hat{G}_{\mathbf{k}}(i\omega),$
where $\hat{G}_{\mathbf{k}}(i\omega)=[i\omega-\hat{\mathcal{H}}_{\text{BdG}}(\mathbf{k},\mathbf{A})]^{-1}$
is the Green's function. 

In leading order in the vector potential, the diamagnetic response
in the Coulomb gauge is given by $\langle j_{i}\rangle=K_{ij}A_{j},$
where $K_{ij}$ is the London kernel. For anisotropic superconductors
that preserve inversion symmetry, the kernel has the form $K_{ij}=(\delta_{ij}-\hat{k}_{i}\hat{k}_{j})Q_{j}$,
with $\hat{k}$ a unitary vector \cite{SM}. The off-diagonal components
of the Kernel result from phase modes \cite{Millis}, which ensure
that the static continuity equation \textbf{$\mathbf{k}\cdot\langle\mathbf{j}\rangle=0$}
is satisfied in the Coulomb gauge. Alternatively, we can simply fix
the gauge in such a way that $\langle j_{i}\rangle=Q_{i}A_{i}$. After
a proper regularization of the deep energy states at the bottom of
the band \cite{Uchoa3}, what is done by imposing periodic boundary
conditions at the edge of the Brillouin zone \cite{SM}, the London
kernel per node is
\begin{equation}
Q_{i}=\frac{e^{2}}{\hbar^{2}c}\Delta^{2}\sum_{\mathbf{p}}\partial_{E_{\mathbf{p}}}\left[\frac{\tanh(E_{\mathbf{p}}/2T)}{E_{\mathbf{p}}}\right]\frac{[\partial_{k_{i}}\mathbf{h}(\mathbf{p})]^{2}}{E_{\mathbf{p}}},\label{eq:Q}
\end{equation}
restoring $\hbar$. Although $\langle j_{i}\rangle$ is calculated
in a fixed gauge, gauge invariance is restored by screening effects
\cite{note}, which preserve the transversality condition of the supercurrent,\textbf{
$\mathbf{k}\cdot\langle\mathbf{j}\rangle=0$}, irrespective of the
gauge choice \cite{Pines}. 

In the semi-Dirac case, the supercurrent due to each node is anisotropic,
as expected, with
\begin{equation}
\langle j_{x}\rangle(T)=\frac{e^{2}}{\hbar^{2}c}\frac{\Gamma^{2}\left(\frac{3}{4}\right)}{\pi^{5/2}}\Theta_{1}(T)\frac{1}{\sqrt{m}v}A_{x},\label{eq:jx}
\end{equation}
and 
\begin{equation}
\langle j_{y}\rangle(T)=\frac{e^{2}}{\hbar^{2}c}\frac{K\left(\frac{1}{2}\right)}{2\pi^{2}}\Theta_{0}(T)\sqrt{m}vA_{y},\label{eq:jy}
\end{equation}
where $\Theta_{n}(T)=\Delta^{2}\!\int_{0}^{\infty}\mbox{d}h\,h^{n}\sqrt{h}\frac{1}{E}\partial_{E}\left[\tanh\!\left(\frac{E}{2T}\right)/E\right]$,
with $E=\sqrt{h^{2}+\Delta^{2}}$. This integral can be analytically
calculated in the zero temperature limit and close to the critical
temperature, 
\begin{equation}
\Theta_{0}(T)=-\begin{cases}
\frac{1}{\sqrt{\pi}}\Gamma^{2}(\frac{3}{4})\sqrt{\Delta} & \mbox{, for }T=0\\
a_{0}\frac{\Delta^{2}(T)}{(2T)^{\frac{3}{2}}} & \mbox{, for }T\approx T_{c}
\end{cases}\label{eq:Theta0}
\end{equation}
where $a_{0}=\int_{0}^{\infty}\mbox{d}x\,x^{-\frac{3}{2}}[x^{-1}\tanh x-\mbox{sech}^{2}x]\approx0.79$,
and 
\begin{equation}
\Theta_{1}(T)=-\begin{cases}
\frac{1}{4\sqrt{\pi}}\Gamma^{2}(\frac{1}{4})\Delta^{\frac{3}{2}} & \mbox{, for }T=0\\
a_{1}\frac{\Delta^{2}(T)}{\sqrt{2T}} & \mbox{, for }T\approx T_{c}.
\end{cases}\label{eq:Theta1}
\end{equation}
with $a_{1}=\frac{1}{2}\int_{0}^{\infty}\mbox{d}x\,x^{-\frac{3}{2}}\tanh x\approx1.91$. 

Near the critical temperature, the kernel anisotropy $\delta(T,g)\equiv Q_{x}/Q_{y}\propto T_{c}(g)/(mv^{2})$
scales linearly with $T_{c}$ and vanishes at the QCP. In the zero
temperature limit, $\delta(0,g)\sim\Delta(g)/(mv^{2})$ and hence
the anisotropy $\delta(0,g)\to0$ lineraly with the gap as one approaches
the QCP at $g=g_{c}$ (orange line in Fig 3). In that limit, the system
is extremely anisotropic \cite{note 2}, with relativistic quasiparticles
carrying a supercurrent along the direction of linear dispersion.
In Fig. 3, we show the plot of the anisotropy per node $\delta$ versus
the gap $\Delta(T_{0},g)$ for fixed temperatures $T_{0}$. When $\Delta\lesssim T_{0}$
, the kernel $Q_{i}$ has a crossover from the anomalous zero temperature
scaling regime, $Q_{x}\propto\Delta^{\frac{3}{2}}$, $Q_{y}\propto\sqrt{\Delta}$,
to the standard BCS scaling, $Q_{i}\propto\Delta^{2}$, where the
anisotropy $\delta(T_{0},g)$ saturates to a constant. 

\begin{figure}
\begin{centering}
\includegraphics[scale=0.45]{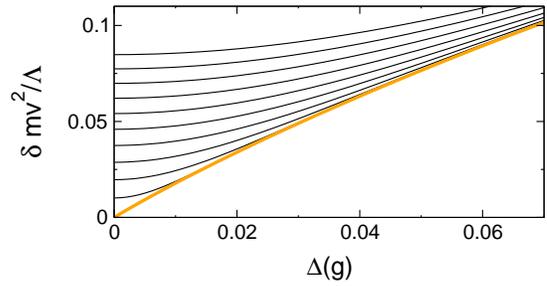}
\par\end{centering}
\caption{{\small{}(color online) Anisotropy $\delta\equiv Q_{x}/Q_{y}$ per
node times $mv^{2}/\Lambda$ versus coupling $\Delta(T_{0},g)$ (in
$\Lambda$ cut-off units) for different temperatures $T_{0}$. $T_{0}/\Lambda$
ranges from zero (orange line) to $0.025$ in $0.0025$ steps. At
$T_{0}=0$, $\delta$ scales to zero at the QCP. In that limit, the
Meissner response becomes quasi-one dimensional.}}
\end{figure}

\emph{Quantum fluctuations$.-$}Allowing the condensate to flow with
momentum $\mathbf{k}_{s}=(k_{x},k_{y})$, we expand the free energy
at zero temperature in powers of the order parameter $\phi$ and $\mathbf{k}_{s}$.
The Ginzburg-Landau (GL) free energy, which fully includes fluctuation
effects, is
\begin{equation}
F_{\text{GL}}=\left(\frac{c_{x}k_{x}^{2}}{\sqrt{m}v}|\phi|^{\frac{3}{2}}+c_{y}\sqrt{m}vk_{y}^{2}\sqrt{|\phi|}\right)+r(g)\phi^{2}+u|\phi|^{\frac{5}{2}},\label{eq:FGL}
\end{equation}
where $\phi=\Delta+\delta\phi$ gives the order parameter around the
saddle point solution $\Delta$ in Eq. (\ref{eq:Delta}), $r(g)=(g^{-1}-g_{c}^{-1})$,
$u=\frac{4}{5}c_{1}\rho_{0}$, $c_{x}=N_{0}\Gamma^{2}\left(\frac{3}{4}\right)\Gamma^{2}\left(\frac{1}{4}\right)/32\pi^{3}$
and $c_{y}=N_{0}K\left(\frac{1}{2}\right)\Gamma^{2}\left(\frac{3}{4}\right)/16\pi^{\frac{5}{2}}$. 

At finite magnetic field, \textbf{$\mathbf{k}_{s}=-(2e/\hbar c)\mathbf{A}$
}by\textbf{ }a suitable gauge choice. Near the QCP, the GL supercurrent
$\mathbf{j}_{s}=c\partial F_{\text{GL}}/\partial\mathbf{A}$ independently
recovers Eq. (\ref{eq:jx}) and (\ref{eq:jy}) at $T=0$. Hence, the
anisotropic quantum critical scaling of the London kernel with $\phi$,
namely $Q_{y}\propto\sqrt{|\phi|}$ and $Q_{x}\propto|\phi|^{\frac{3}{2}}$,
persists near the QCP, where quantum fluctuations dominate. 

Because the free energy (\ref{eq:FGL}) has non-analytic terms both
in the kinetic energy and in the interaction term $u$, one cannot
expand in the fluctuation fields $\delta\phi$ in order to integrate
them out and calculate the quantum fluctuation corrections to the
scaling of $\Delta(0,g)\propto(g-g_{c})^{\beta}$, with $\beta=2$
in mean field \cite{note3}. Instead, one needs to resort to field
theoretical methods \cite{Vojta,Khveschenko}, which are beyond the
scope of this work and will be addressed elsewhere. In any case, the
mean field analysis is accurate in the regime where the quadratic
term of (\ref{eq:FGL}) dominates over the interaction term $u$,
namely $(g/g_{c}-1)^{2}\gtrsim N_{0}^{-1}v/(\sqrt{m}\Lambda^{\frac{3}{2}})$.

\emph{Penetration depth.}$-$ For a thin film of thickness $d$, the
penetration depth is given by the London kernel, $\lambda_{i}=\sqrt{-cd/(4\pi Q_{i})},$
with $i=x,\,y$. In general, for systems of semi-Dirac fermions with
uniaxial anisotropy, such as in uniaxially strained graphene or semi-metallic
black phosphorus, the total London kernel is calculated from the Meissner
response of a single nodal point times the nodal degeneracy $N_{0}$.
In that case, at zero temperature,
\begin{equation}
\lambda_{x}\propto\frac{\hbar c}{e}\sqrt{d}\,\Delta^{-\frac{3}{4}}(g)\left(\sqrt{m}v/N_{0}\right)^{\frac{1}{2}},\label{eq:lambda}
\end{equation}
 and
\begin{equation}
\lambda_{y}\propto\frac{\hbar c}{e}\sqrt{d}\,\Delta^{-\frac{1}{4}}(g)/\left(\sqrt{m}vN_{0}\right)^{\frac{1}{2}},\label{eq:lambda2}
\end{equation}
and hence the penetration depth along the $x$ and $y$ axes grows
near the QCP with different scaling exponents, $\lambda_{x}(g)\propto(1-g_{c}/g)^{-3\beta/4}$
and $\lambda_{y}(g)\propto(1-g_{c}/g)^{-\beta/4}$. Near the critical
temperature, the penetration depth is still anisotropic, but follows
the standard BCS temperature scaling $\lambda\propto\Delta^{-1}(T)$. 

\emph{Coherence length}.$-$ In the zero temperature limit, the coherence
length $\xi_{0}$ corresponds to the length scale where the energy
of the system changes by an amount set by the mass gap $2\Delta$.
Near the neutrality point ($\mu\ll\Delta$, with $\mu$ the chemical
potential away from half filling), the corresponding change in the
momentum domain $\delta p$ satisfies $h(\delta p)\sim2\Delta$. Since
$\xi_{0}\sim\hbar/\delta p$, variations along the direction where
the energy spectrum is linear imply that $\xi_{0,y}\sim\hbar v_{y}/(2\Delta)$.
A similar dimensional analysis along the direction of parabolic dispersion
gives $\delta p_{x}\sim\sqrt{2m\Delta}$, and hence 
\begin{equation}
\xi_{0,x}\sim\hbar/\sqrt{2m\Delta},\label{eq:xi}
\end{equation}
in contrast with the standard Fermi liquid result ($\mu\gg\Delta$),
where $\xi_{0}\equiv\hbar v_{F}/(\pi\Delta)$, with $v_{F}$ the Fermi
velocity \cite{Tinkham}. Fluctuation effects are expected to give
small deviations in the quantum critical scaling of the coherence
length with $\Delta$ due to the emergence of an anomalous dimension. 

In mean field, the ratio between the penetration depth in the London
limit and the coherence length $\kappa=\lambda/\xi_{0}$ is given
by
\begin{equation}
\kappa_{x}\sim\Delta^{-\frac{1}{4}}(g)\left(\sqrt{m}v\right)^{\frac{1}{2}}c\frac{\sqrt{md}}{e},\label{eq:kx}
\end{equation}
and
\begin{equation}
\kappa_{y}\sim\Delta^{\frac{3}{4}}(g)\left(\sqrt{m}v\right)^{-\frac{1}{2}}\frac{c}{v}\frac{\sqrt{d}}{e}\label{eq:ky}
\end{equation}
along the two principal directions $x$ and $y$, with proportionality
factors of the order of 1. Therefore, in the vicinity of the QCP,
the order parameter becomes rigid for amplitude variations along the
direction where the quasiparticles have linear dispersion ($\kappa_{y}\propto(1-g_{c}/g)^{3/2}\ll1$),
as in type $I$ superconductors. At the same time, the order parameter
becomes soft for variations along the direction of parabolic dispersion
($\kappa_{x}\propto(1-g_{c}/g)^{-1/2}\gg1$), as in type II superconductors.
While fluctuations could provide corrections to the scaling of $\kappa$,
the mean field analysis is suggestive of a possible smetic instability
near the QCP.

\emph{Stripe phase.$-$}The energy of a domain wall becomes negative
when $\kappa>1/\sqrt{2}$. Near the QCP, the magnetic flux can form
a stripe pattern of domain walls oriented along the $y$ direction,
which coincides with the ``easy'' direction for the supercurrent
as indicated in Fig. 4a. Those domains separate superconducting regions
(S), which are screened by diamagnetic currents (red arrows in Fig.
4a), from normal regions (N) of width $\sim2\xi_{0,x}$ separated
by a distance $l\gg\xi_{0,x}$. Because the magnetic field $H$ has
a stiffness of the order of the penetration depth $\lambda_{x}\gg\xi_{0,x}$
along the $x$ direction, those domain walls of magnetic flux repel
each other and can stabilize a stripe phase in the regime where the
magnetic field normal to the sample is strong enough. 

\begin{figure}[t]
\begin{centering}
\includegraphics[scale=0.42]{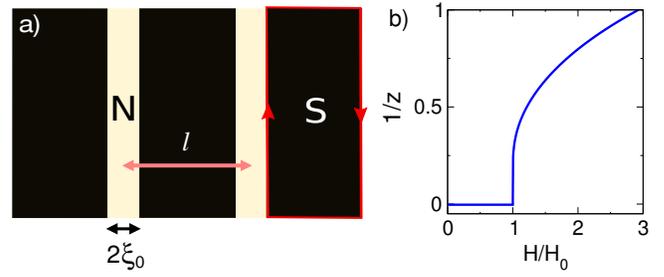}
\par\end{centering}
\caption{{\small{}(color online) a) Stripe phase of superconducting domains
(S) oriented along the direction where the order parameter is stiff.
The normal regions (N) have a magnetic field $H$, and width of twice
the coherence length $\xi_{0}$. The separation between the center
of the stripes is $l\gg\xi_{0}$. Red lines: diamagnetic currents.
b) Scaling of $z=l/\lambda$ versus the magnetic field $H$. For $H\leq H_{0}\equiv H_{c}/\sqrt{\kappa}$,
$l\to\infty$. For $H>H_{0}$, $l$ is finite. }}
\end{figure}

Domain wall formation in the bulk of macroscopic samples is ellusive
and has been observed only in a few ferromagnetic superconductors
\cite{Goldstein,Prozorov,Wang}. For samples with finite slab geometry,
domain walls are observed in the intermediate state of type I superconductors,
where the period of the laminar state is set by the thickness of the
sample, $l\propto\sqrt{d}$. In semi-Dirac metals with uniaxial anisotropy,
the stripe phase will have lower energy compared to the vortex state
of type II superconductors near the QCP. In the presence of magnetic
fields, the Gibbs free energy of a striped normal domain surrounded
by superconducting regions of width $l$ is \cite{DeGennes} 
\begin{equation}
G(H,z)=\frac{1}{8\pi z}\left(\frac{H_{c}^{2}}{\kappa_{x}}-H^{2}\tanh z\right),\label{eq:G}
\end{equation}
where $z=l/\lambda_{x}$ is the distance between the normal domain
walls normalized by the penetration depth and $H_{c}$ is the field
that corresponds to the condensation energy $H_{c}^{2}/8\pi$. The
equilibrium separation between the stripes follows trivially from
minimization of the free energy for fixed field, $\partial G(H,z)/\partial z=0$. 

In Fig. 4b, we show the scaling of $z=l/\lambda_{x}$ as a function
of the magnetic field $H$. Below the critical field $H<H_{c}/\sqrt{\kappa_{x}}$,
$l\to\infty$, and the system has a uniform phase (Meissner state).
In the regime $H_{c}/\sqrt{\kappa_{x}}<H\lesssim H_{c}\kappa_{x}$
, $l$ is finite and the system will form a smetic state with stripes
of superconducting domains separated by thin strips of magnetic flux.
Eventually, when $H\gtrsim H_{c}\kappa_{x}$, the separation of the
domains $l\sim\xi_{0,x}$ \cite{DeGennes} and superconductivity will
be destroyed. 

\emph{Conclusions.$-$ }In summary, we examined the critical properties
of semi-Dirac metal superconductors at zero and finite magnetic fields.
We showed that near the quantum critical regime and at finite fields,
the anisotropy of the quasiparticles leads to an exotic electromagnetic
response which may stabilize a novel smetic state of superconducting
stripes. 

\emph{Acknowledgements.$-$} BU thanks K. Mullen, M. Fogler and S.
Parmeswaran for helpful discussions. BU acknowledges NSF CAREER grant
DMR-1352604 for support.


\begin{thebibliography}{10}
\bibitem{Banarjee}S. Banerjee, R. R. P. Singh, V. Pardo, and W. E.
Pickett, Phys. Rev. Lett. \textbf{103}, 016402 (2009). 

\bibitem{Young}S. M. Young, and C. L. Kane, Phys. Rev. Lett. \textbf{115},
126803 (2015).

\bibitem{Huang}H. Huang, Z. Liu, H. Zhang, W. Duan, and D. Vanderbilt,
Phys. Rev. B \textbf{92}, 161115(R) (2015). 

\bibitem{Saha}K. Saha, Phys. Rev. B \textbf{94}, 08113(R) (2016).

\bibitem{Pardo}V. Pardo and W. E. Pickett, Phys. Rev. Lett. \textbf{102},
166803 (2009).

\bibitem{Montambaux}G. Montambaux, F. Piéchon, J.-N. Fuchs, and M.
O. Goerbig, Phys. Rev. B \textbf{80}, 153412(2009).

\bibitem{Rodian}A.\LyXThinSpace S. Rodin, A. Carvalho, and A.\LyXThinSpace H.
Castro Neto, Phys. Rev. Lett. \textbf{112}, 176801 (2014).

\bibitem{Kim}J. Kim \emph{et al., }Science\emph{ }\textbf{349}\emph{,
}723 (2015).

\bibitem{note3-1}This state is fundamentally different from multi-component
``type-1.5'' superconductivity, where competing coherence lengths
lead to phase separation into \emph{vortex} stripes and clusters.
See V. Moshchalkov \emph{et. al.}, Phys. Rev. Lett. \textbf{102},
117001 (2009). 

\bibitem{Adroguer}P. Adroguer, D. Carpentier, G. Montambaux, and
E. Orignac, Phys. Rev B \textbf{93}, 125113 (2016).

\bibitem{Kotov}V. N. Kotov, B. Uchoa, V. M. Pereira, F. Guinea, and
A. H. Castro Neto, Rev. Mod. Phys. \textbf{84}, 1067 (2012). 

\bibitem{Uchoa}B. Uchoa, and A. H. Castro Neto, Phys. Rev. Lett.
\textbf{98}, 146801 (2007).

\bibitem{Zhao}E. Zhao and A. Paramekanti, Phys. Rev. Lett. \textbf{97},
230404 (2006).

\bibitem{Wei}Black phosphorus can sustain strain deformations of
$30\%$, which could lead to a comparable reduction in the velocity
and enhancement in the mass of the quasiparticles. See Q. Wei and
X. Peng, Appl. Phys. Lett. \textbf{104}, 251915 (2014). 

\bibitem{deltaC}The normalized specific heat jump is $\delta C_{V}=\sqrt{2}\gamma_{0}^{2}\pi^{3/2}/[\gamma_{2}(4\sqrt{2}-1)\zeta\!\left(\frac{5}{2}\right)]\approx0.71$
where $\zeta\left(\frac{5}{2}\right)\approx1.34$ is a zeta function,
and $\gamma_{n}\equiv\int_{0}^{\infty}\mbox{d}x\,x^{n}\sqrt{x}\,\mbox{sech}^{2}x$,
with $\gamma_{0}\approx0.76$ and $\gamma_{2}\approx1.02$. 

\bibitem{uchoa}B. Uchoa, G.G. Cabrera, and A. H. Castro Neto, Phys.
Rev. B \textbf{71}, 184509 (2005). 

\bibitem{Tinkham}M. Tinkham, Introduction to superconductivity, Dover,
1996. 

\bibitem{SM}See supplementary materials. 

\bibitem{Millis}A. J. Millis, Phys. Rev. B \textbf{35}, 151 (1987). 

\bibitem{Uchoa3}B. Uchoa, and A. H. Castro Neto, Phys. Rev. Lett.
\textbf{102}, 109701 (2009).

\bibitem{note}At long wavelengths, virtual plasmons screen the longitudinal
component of the supercurrent, mantaining the response of the system
invariant under any gauge (see Ref. \cite{Pines}). For a slab of
thickness $d$, optical plasmons provide screening in the $q\to0$
limit in the presence of any arbitrarily small pocket of charge with
energy $\mu$ around the nodal points. When $\mu\ll T\ll\Delta$,
quantum criticality is reminiscent and drives the critical scaling
of the physical observables. 

\bibitem{Pines}D. Pines and J. R. Schrieffer, Nuovo Cimento \textbf{10},
496 (1958). 

\bibitem{note 2}In Hamiltonian (1), the four-fold rotational symmetry
of the lattice is restored by the second pair of nodes along the $(1,\bar{1})$
direction of the crystal. 

\bibitem{note3}For Dirac fermions, $\beta=1$ at the mean field level
and $\beta\approx0.877$ according to quantum Monte Carlo calculations.
See L. Karkkainen, R. Lacaze, P. Lacock, and B. Petersson, Nucl. Phys.
B\textbf{ 415 }781 (1994). 

\bibitem{Vojta}M. Vojta, Y. Zhang, S. Sachdev, Int. J. Mod. Phys.
B \textbf{14}, 3719 (2000). 

\bibitem{Khveschenko}D. V. Khveschenko, J. Paaske, Phys. Rev. Lett.
\textbf{86}, 4672 (2001). 

\bibitem{DeGennes}P. G. DeGennes, Superconductivity of metals and
alloys, Addison Wesley, 1989.

\bibitem{Goldstein}R. E. Goldstein, D. P. Jackson, and A. T. Dorsey,
Phys. Rev. Lett. \textbf{76}, 3818 (1996).

\bibitem{Prozorov}R. Prozorov, Phys. Rev. Lett. \textbf{98}, 257001
(2007).

\bibitem{Wang}X. Wang, M. Mostovoy, M. G. Han, Y. Horibe, T. Aoki,
Y. Zhu, and S.-W. Cheong, Phys. Rev. Lett. \textbf{112}, 247601 (2014).
\end{thebibliography}
\end{document}